# Ultrahigh-Fidelity Spatial Mode Quantum Gates in High-Dimensional Space by Diffractive Deep Neural Networks

Qianke Wang[1,2,†], Jun Liu[1,2,†], Dawei Lyu[1,2], Jian Wang[1,2,*]

[1] *Wuhan National Laboratory for Optoelectronics and School of Optical and Electronic Information, Huazhong University of Science and Technology, Wuhan 430074, Hubei, China*

[2] *Optics Valley Laboratory, Wuhan 430074, Hubei, China*

[†] *These authors contributed equally to this work.*

[*] *Corresponding Author: jwang@hust.edu.cn*

# Abstract

While the spatial mode of photons is widely used in quantum cryptography, its potential for quantum computation remains largely unexplored. Here, we showcase the use of the multi-dimensional spatial mode of photons to construct a series of high-dimensional quantum gates, achieved through the use of diffractive deep neural networks ($D^2$NNs). Notably, our gates demonstrate high fidelity of up to 99.6(2)%, as characterized by quantum process tomography. Our experimental implementation of these gates involves a programmable array of phase layers in a compact and scalable device, capable of performing complex operations or even quantum circuits. We also demonstrate the efficacy of the $D^2$NN gates by successfully implementing the Deutsch algorithm and propose an intelligent deployment protocol that involves self-configuration and self-optimization. Moreover, we conduct a comparative analysis of the $D^2$NN gate's performance to the wave-front matching approach. Overall, our work opens a door for designing specific quantum gates using deep learning, with the potential for reliable execution of quantum computation.

# Introduction

Quantum logic gates are a fundamental component of quantum information processing[1] and reliable quantum computing necessitates quantum operations with high fidelity. To date, quantum gates based on superconducting circuits[2,3], trapped ions[4,5], defects in solid states[6] and spin qubit in silicon[7–9] have met the error correction threshold of 99% in surface code[10]. However, similar levels of performance have not yet been observed in photonic quantum logic gates[11–17], which possess unique features that have attracted considerable interest in the field of quantum information processing. Photonic quantum gates are naturally compatible with quantum communications, which result in excellent networkability[18–23]. Benefiting from their weak interactions with the environment, photons often have long decoherence time, low noise and crosstalk, thus demanding fewer overheads for fault-tolerance[24]. Moreover, photons are low-cost quantum systems that operate at room temperature compared to other quantum systems. By exploiting their abundant physical dimension resources and modulation flexibility, photonic quantum gates have been demonstrated in many degrees of freedom (DoFs)[25–28].

Among all these DoFs, the intrinsic infinity of orthogonal bases in the spatial modes of photons offers an extensive coding alphabet, encouraging creativity in high-dimensional quantum information processing. High-dimensional quantum states increase the information per photon and exhibit robustness against environmental noise[29] and quantum cloning. The realization of quantum gates on the spatial modes of photons has garnered significant attention due to these advantages. Discrete optics schemes have already demonstrated the quantum *controlled-NOT* (*CNOT*) gate[11], four-dimensional *X* gate[13], four-dimensional *CNOT* gate[14], and *controlled-SWAP* gate[15]. However, despite these excellent works, achieving high-dimensional unitary transformations in a scalable and compact manner with high fidelities remains a challenge.

Recently, researchers have attempted to achieve coaxial unitary transformations through multi-plane light-conversion with impressive performance[16]. However, this approach lacks process tomography fidelities and requires further research. In contrast to the traditional method of wave-front matching (WFM)[30], more sophisticated approaches based on machine learning have been developed[31]. One such approach, the all-optical machine learning framework known as diffractive deep neural networks ($D^2$NNs), was originally proposed for optical processing tasks such as image recognition and classification, with low energy costs and at the speed of light[32]. Many variants of the original design have since been reported to improve its performance and expand its applications[33–36]. Recently, it has been shown that $D^2$NNs are capable of performing arbitrary complex-valued linear transformations[37].

Here we propose a novel scheme for implementing a variety of quantum gates using the deep-learning-based design of passive diffractive layers placed in sequence to manipulate the spatial modes of photons. To be specific, we demonstrate three-dimensional $X$ gates and $H$ gates, as well as a single-photon $CNOT$ gate through the adoption of specialized coding rules, leading to remarkable process fidelities exceeding 99%. Our compact and reconfigurable implementation displays great scalability and robustness of mode basis and is conducive to intelligent deployment. Moreover, we demonstrate the applicability of our approach by employing the Deutsch algorithm, which validates the feasibility of constructing a spatial mode quantum computer.

## Results

**Concept of quantum gate assisted by $D^2$NN architecture**

Fig. 1a illustrates the quantum gate implementation concept. The spatial mode quantum gate is responsible for converting the spatial modes, where different mapping rules correspond to different logical gate operations. To design a specific quantum gate, the input and output mode correspondence should be determined. As an example, we use three Laguerre-Gaussian modes ($LG_0^{-2}$, $LG_1^0$ and $LG_0^2$) as inputs and outputs for the neat axial symmetry of their superposition states and exploiting azimuthal and radial DoFs to show the mapping of the three-dimensional *X* gate. It should be clarified that the notion of $LG_p^\ell$ represents the LG mode with azimuthal order of $\ell$ and radial order of $p$. The mode phases and the mapping between them are represented on the side of the D²NN for clarity, with the yellow line indicating the current working mode conversion in this concept figure. To implement this gate, we need to generate phase layers, which are where the D²NN comes into play. The D²NN, inspired by digital artificial neural networks, uses phase planes as hidden layers to learn the relation between inputs and outputs in an energy- and time-efficient way by harnessing the power of light. In the D²NN, inputs and outputs are optical fields, and the hidden layers are represented by phase planes, whose pixel phase determines the layer weight. Pixels between two layers are linked by the diffraction effect, and multiple phase layers are aligned sequentially along the optical axis to accomplish the desired mode conversion. The model of the spatial mode quantum gate is built and ready to be trained by iterative algorism on a computer. To learn the optimal phase layers that implement the desired operation on the experiment setup, we minimize the loss function of the deviation between the inference output field and the theoretical output field by the Adam gradient-descent algorithm[31] while the pixel phase values are updated in diffractive layers. The optimization result is obtained when the loss function converges. See Supplementary Note 1 for more details on the pattern generation.

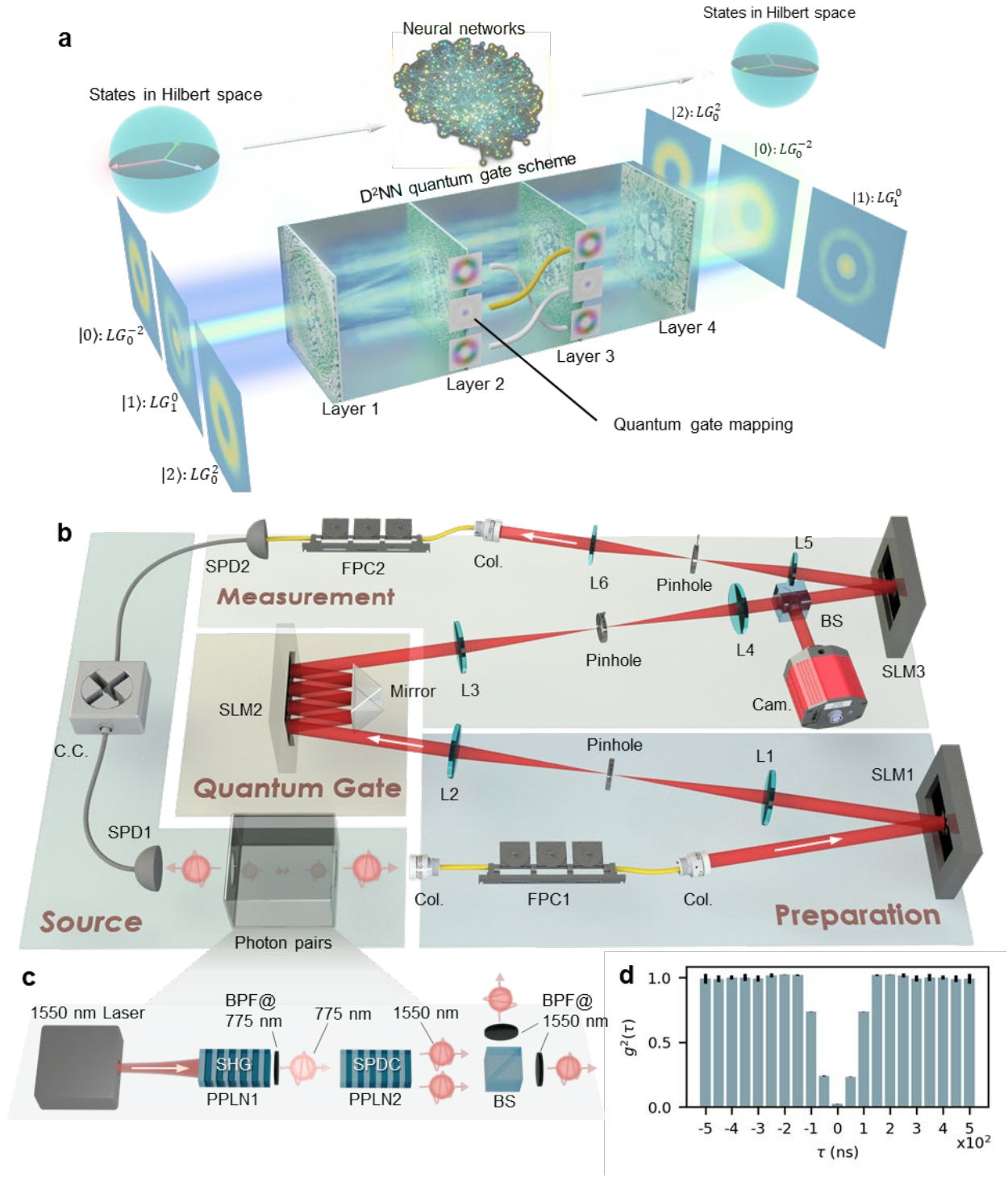

**Fig. 1 Architecture of the spatial mode quantum gate using diffractive deep neural networks (D²NNs).**
**a** Conceptual illustration of the D²NN quantum gate's abstract architecture, showing how the neural networks map quantum states in high-dimensional Hilbert space to other states in that space. The specific three-dimensional $X$ gate scheme is shown in the lower part as an example, including input states, multilayer D²NN, and output states. The mapping between inputs and outputs is indicated on the side of the D²NN, with the currently active mapping highlighted in yellow. **b** Experimental setup of a 4-layers spatial mode quantum gate. **c** A heralded single photon source is used. FPC: fiber polarization controller; Col.: collimator; SLM: spatial light modulator; L1-L6: lenses; BS: beam splitter; Cam.: camera; SPD1&2: single photon detector; C.C.: coincidence counting; PPLN1&2: periodically poled lithium niobate; SHG: second harmonic generation; SPDC: spontaneous parametric down-conversion; BPF: bandpass filter.

While the heralding photons of the photon pairs are sent to the SPD1, the heralded photons are prepared into input states by complex modulation using SLM1, followed by D²NN-generated phase layers loaded onto SLM2 to perform the desired operation. Output states are analyzed using SLM3, a fiber coupling system, and the SPD2. The coincidence counting of SPD1&2 indicates the single photon passing the entire system. **d** Second-order correlation function $g^2(\tau)$ of the heralded single photon source determined using the Hanbury Brown and Twiss setup. The $g^2(0) = 0.024(2)$ implies the remarkable performance of the source.

To verify the functionality of the generated phase layers, we conduct an experiment in which we imprint them onto a spatial light modulator (SLM), as depicted in Fig. 1b. The device comprises four parts: source, preparation, quantum gate, and measurement. The source is made by a heralded single photon source which detects one photon (known as heralding photon) from a photon pair to announce the arrival of the other one (the heralded photon), as show in Fig. 1c. Here the photon pairs at 1550 nm are generated by spontaneous parametric down-conversion (SPDC) in a type-0 periodically poled lithium niobite bulk crystal (PPLN2). The pump beam at 775 nm are generated via second harmonic generation of PPLN1 (see Method for more details). The second-order correlation function $g^2(\tau)$ of the heralded single photon source, characterized through the Hanbury Brown and Twiss (HBT) setup, is illustrated in Fig. 1d. The value of $g^2(0)$ is measured to be 0.024(2) based on the experimental outcomes derived from three-fold coincidences, with delays being varied. When assessed using another commonly employed evaluation method, $g^2(0) = C_{123} \times N_3 / C_{13} / C_{23}$, the value of $g^2(0)$ could be as low as $2.56 \times 10^{-4}$. Here, $C_{123}$ is the coincidence count of channel 1, 2 and 3 in the HBT setup, $N_3$ is the single count of channel 3, and $C_{13}(C_{23})$ denotes the coincidence count of channels 1 and 3 (2 and 3).

The heralded photons are then guided to the preparation stage, where we employ the complex modulation technique[38,39] to prepare the desired states and to project output states, and load SLM2 with the generated D²NN phases to implement the quantum gate operation. To optimize the efficiency of SLM usage, simplify the experimental setup, and reduce the footprint of the device, we set a mirror opposing the SLM2 and adjust the incident angle such that the light reflects on the SLM2 the same number of times as the number of layers. We note that the incident angle will affect the alignment since layers are perpendicular to the optical axis in the modeling described in Fig. 1a. But the degradation in this experiment is not significant and thus acceptable here. Further analysis of the effect of the incident angle can be found in ref.[40].

**High-dimensional spatial mode gates, single-photon CNOT gate and process tomography**

The high-dimensional spatial mode gate performs a crucial function in higher dimensional encoding space. We design and demonstrate three-dimensional $X$ gates, $H$ gates, and a single-photon $CNOT$ gate, as the diagrams depicted in Fig. 2a-c. The three-dimensional $X_1$ gate circularly shifts the basis states forward, while the three-dimensional $H_1$ gate transforms the basis states into superpositions of all basis states with different well-defined phases. For the $CNOT$ gate, we adopted a unique coding method to encode two bits of information, utilizing four orbital angular momentum (OAM) modes of a single photon, which are $|-1\rangle$, $|+1\rangle$, $|-3\rangle$ and $|+3\rangle$ corresponding to $|00\rangle$, $|01\rangle$, $|10\rangle$ and $|11\rangle$. With this method, input state $|-3\rangle$ is flipped to $|+3\rangle$ by this $CNOT$ gate and vice versa, while input states $|\pm1\rangle$ remaining unchanged. This method is inspired by the concept of path coding. Consider the scenario where four distinct path states are at disposal; they can be interpreted as a tensor product of two two-dimensional qubits, such as the top/bottom and left/right directions, as depicted in the upper section of Fig. 2d. Similarly, the OAM DoF, also possessing infinite dimensions, can be viewed as the "path" within the mode space. We artificially categorize four OAM states into two types: sign dimensions (referring to phase rotation directions) and order dimensions (indicating phase rotation orders). Each category comprises two levels, effectively forming the dual levels for control and target qubits. This configuration is illustrated in the lower segment of Fig. 2d. The path encoding method has been employed for demonstrating quantum fault-tolerant threshold recently[41]. Additionally, the mode encoding method has also been utilized, with reference to the azimuthal and radial order of LG modes[16]. Fig. 2e-g display the output mode profiles of these gates, which exhibit a high degree of consistency between the simulations and experimental measurements. The minor deviations in the output modes might be attributed to possible misalignment and imperfections in the apparatus. The corresponding inputs are prepared in all mutually unbiased bases (MUBs) in three-dimension for *X* and *H* gates, while for the *CNOT* gate, an overcomplete state set in 2×2-dimension is utilized. Further details about the MUBs can be found in Supplementary Note 2.

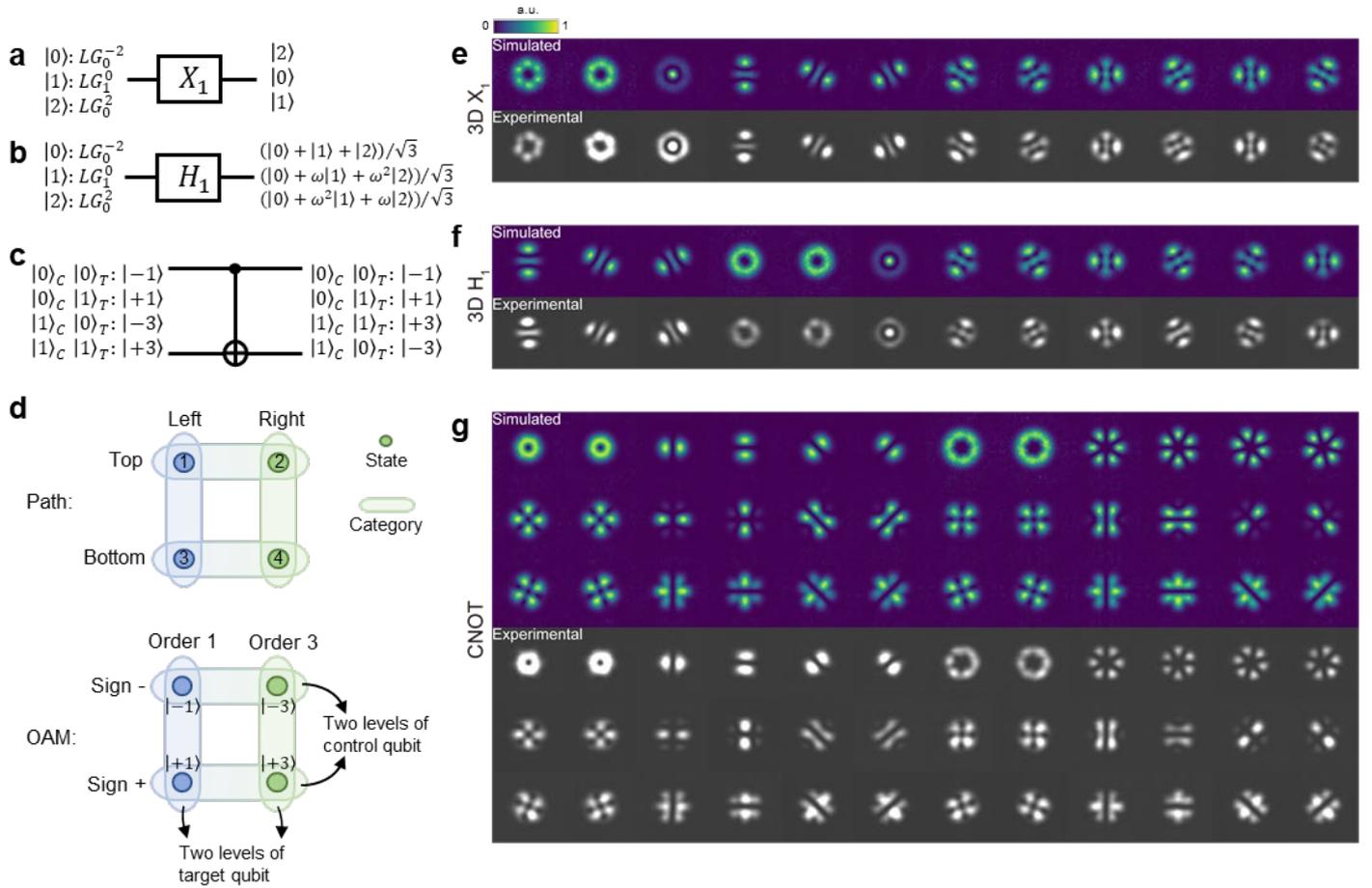

**Fig. 2 High-dimensional spatial mode gates and output modes. a-c** Gate diagrams for the three-dimensional $X_1$ gate, three-dimensional $H_1$ gate and *CNOT* gate, with their input and output states. The phase parameter $\omega$ is $e^{2\pi i/3}$, and the $|0\rangle_C$ ($|0\rangle_T$) denotes that the control (target) qubit is in $|0\rangle$ state.

**d** Comparison between the path encoding and orbital angular momentum (OAM) encoding method. Four distinct states are categorized and can be interpreted as a tensor product of two two-dimensional qubits, such as the top/bottom and left/right directions for the path encoding, and sign dimensions (referring to phase rotation directions) and order dimensions (indicating phase rotation orders) for the OAM encoding.

**e-g** Simulated output mode profiles and experimental profiles captured by a camera, corresponding to input states spanning all mutually unbiased bases (MUBs) in three dimensions for *X* and *H* gates, and in a 2x2 dimension for the *CNOT* gate.

To obtain a more comprehensive characterization of these gates, it is necessary to employ quantum process tomography (QPT)[42]. It involves preparing the input states in all MUBs as previously described, applying the gates to this set of states, and performing full state tomography on every output state by means of projective measurements in all MUBs. The resulting normalized measurement outcomes yield the tomography matrices presented in Fig. 3a-c. With these tomography results, we can infer the quantum process $\varepsilon(\rho_{in})$, which can be decomposed as

$$\varepsilon(\rho_{in}) = \sum_{m,n=1}^{d^2} \chi_{mn} E_m \rho_{in} E_n^\dagger \quad (1)$$

where $\rho_{in}$ is the input state of the system, $d$ is the dimension, and $E$ is a set of operators usually being generalized Pauli matrix. We use the Gell–Mann matrices for $d = 3$ and two-qubit generalized Pauli matrices for $d = 4$. The $d^2 \times d^2$ matrix $\chi$ is the process matrix, which contains all information about this quantum process $\varepsilon(\rho_{in})$ and can be reconstructed since all other elements in Eq. (1) is known. Fig. 3d-f exhibit bar plots of the reconstructed process matrices $\chi$. For a clearer view, the bars with negative values are flipped up and presented with pale grey edges. A physical matrix must be positive semidefinite and trace preserving. Due to inherent experimental noise, however, standard QPT might produce an unphysical matrix with negative eigenvalues. Thus, the maximum-likelihood estimation (MLE) method that always yields physically sensible results is devised for the reconstruction[42,43]. By introducing the Lagrange multiplier and constructing an appropriate iteration form, MLE method preserves the positive semi-definiteness and trace normalization of the process matrix (see Supplementary Note 3). The conformity between theory and experiment is evaluated by process fidelity $F = \text{Tr}(\chi_t \chi_e)$, which is the trace of the product of theoretical process matrix $\chi_t$ and experimental process matrix $\chi_e$. In this experiment, we achieve $F_{3DX1} = 98.4(2)\%$, $F_{3DH1} = 99.4(3)\%$ and $F_{CNOT} = 99.6(2)\%$, respectively. To quantify the comparison between theoretical, simulated and experimental tomography matrices, the mean squared errors (MSEs) of them for different gates are exhibited in Fig. 3g. An intuitive observation emerging from the data comparison is that lower MSE signifies better fidelity, and the experimental imperfections significantly reduce the MSE level. Additionally, the process matrices of $X_2$, $H_2$, $H_3$ are listed in Supplementary Note 2. Our results indicate not only the feasibility of high-dimensional operation in spatial modes of photons using D²NN, but also the high performance which is essential to the reliable execution of quantum algorithms.

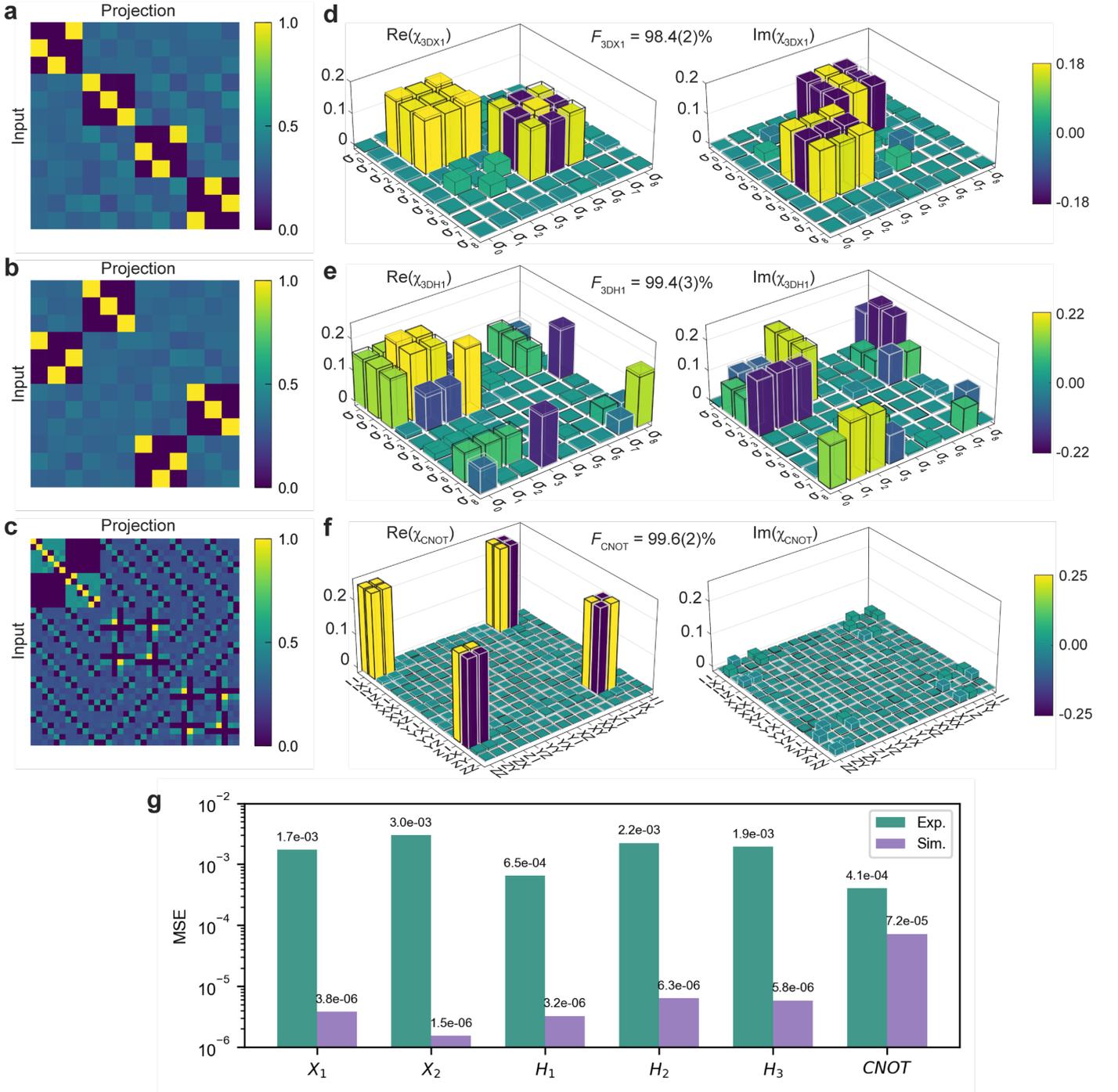

**Fig. 3 Quantum process tomography of spatial mode quantum gates. a-c** Tomography matrices for reconstructing the process matrices of the gates. Each element of the tomography matrix corresponds to the projective measurement outcome for a specific combination of input and projection basis. **d-f** Reconstructed process matrices χ for the three-dimensional $X_1$ gate, three-dimensional $H_1$ gate and CNOT gate. The $X$ and $H$ gates process matrices are shown in the Gell–Mann matrix basis, while the CNOT gate process matrix is shown on the basis of two-qubit generalized Pauli operators. Theoretical process matrices and experimentally reconstructed process matrices are represented by transparent and colored bars, respectively. The real and imaginary parts are plotted separately, and bars with negative values are flipped up for better visibility and have greyish-white edges. The process fidelities for these three gates are 98.4(2)%, 99.4(3)%, and 99.6(2)%. **g** Mean squared error (MSE) of simulated and experimental tomography compared to theoretical tomography matrix.

**Demonstration of the Deutsch algorithm and intelligent deployment**

Quantum gates are fundamental components of quantum algorithms, which leverage quantum superposition and interference to achieve computational advantages over classical algorithms. These gates are capable of simultaneously processing all possible states via superposition and achieving the correct outcome via interference. As an application of the $D^2NN$ quantum gates, we implement the quantum circuit of the two-bit version of the Deutsch-Jozsa algorithm, known as the Deutsch algorithm[44,45]. The algorithm's goal is to determine whether a Boolean function $f:\{x_1,x_2,\cdots,x_n\} \rightarrow \{0,1\}$ is constant ($f(x_i)=0$ or $f(x_i)=1$ for all $x$), or balanced ($f(x_i)=0$ for half of $x$ and $f(x_i)=1$ for the other half of $x$). In classical computing, the worst-case scenario requires $2^{n-1}+1$ query of $f$ to identify it, since there are $2^n$ possible outputs. Leveraging the power of the Deutsch algorithm, the question can be solved with just one function evaluation, instead of growing exponentially with the number of qubits. In the state preparation stage, we set the qubit $x$ as $|0\rangle_x$ and qubit $y$ as $|1\rangle_y$, and apply two $H$ gates to each qubit to create a superposition that represents all possible state combinations. By creating a superposition, the oracle function (maps the state $|x\rangle|y\rangle$ to $|x\rangle|y\oplus f(x)\rangle$, $\oplus$ is the *XOR* operation) can work on all possible configurations simultaneously. After applying the oracle function and the interference (*H* gates to each qubit after the oracle), we will measure qubit $x$ as 0 for all constant functions or 1 for balanced functions. The derivation of this process is in Supplementary Note 4. Two kinds of oracle functions that are constant and balanced are constructed using an identity operator and a *CNOT* operator, respectively, as shown in Fig. 4a. The corresponding output states are presented before measurement. Notably, the oracle function and interference process are implemented with a single SLM, which reduces system complexity and alignment difficulties, enhancing the practicality of our implementation. Although only $x$ qubit measurement is required , we project and measure the output photon in four bases of two qubits for clarity, as shown in Fig. 4b, and the results are consistent with the theoretical output states in Fig. 4a.

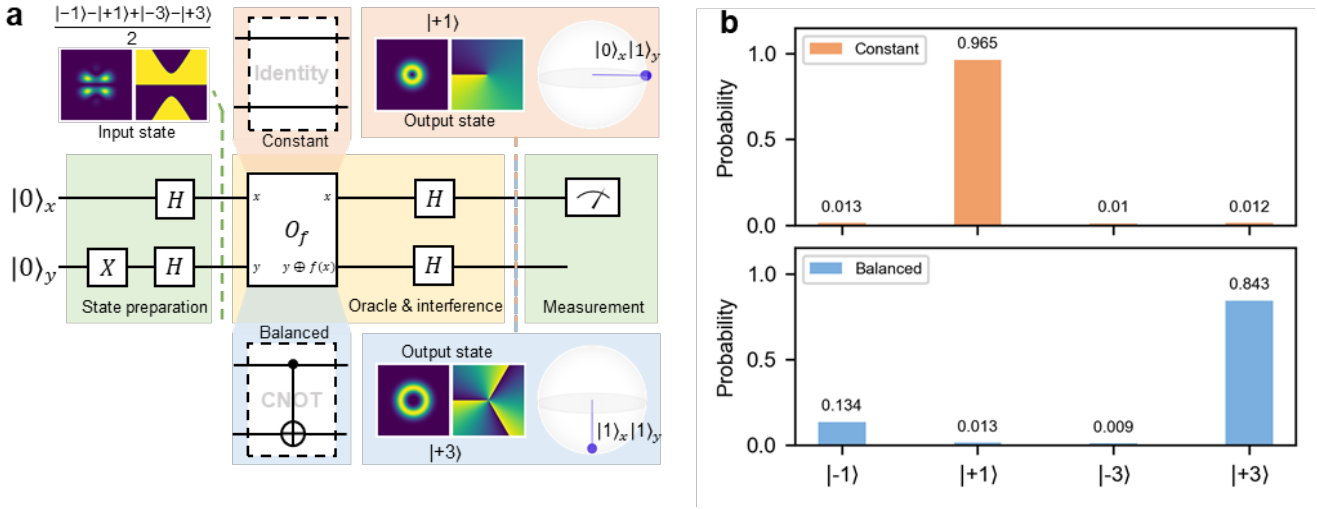

**Fig. 4 Demonstration of the Deutsch algorithm using D²NN spatial mode quantum gates. a** Circuit diagram of the Deutsch algorithm based on the single-photon quantum gate. The whole circuit is divided into three parts: state preparation, oracle operation followed by interference, and measurement, as marked in three colored blocks on the circuit. These three parts are performed by three separate spatial light modulators (SLMs). The oracle is implemented by an identity operation circuit for a constant function and a *CNOT* operation for a balanced function. The states after the first and second parts are illustrated in this figure, separately. **b** Projective measurement results of the circuit output states in four bases for constant (in orange) and balanced (in blue) functions.

In addition to demonstrating various gates in the quantum circuit, our reconfigurable implementation showcases more innovative features. We propose a protocol to explore its potential in applications that demand intelligent deployment. This protocol takes advantage of the flexibility offered by the reconfigurable setup and the smooth coordination among preparation, operation, and measurement devices. The task is to configure the gate setup according to preset commands, specifically achieving the automated switch from $U_1$ to $U_2$. The entire self-configuration process is illustrated in Fig. 5a, encompassing several steps. First, the current gate's performance is assessed using process tomography. This tomography matrix is then utilized for reconstructing the gate fidelity via the MLE method mentioned earlier. The MLE method produces different fidelities based on various assumptions about ideal tomography matrices, as depicted in Fig. 5b. Among these assumptions, the one leading to the highest fidelity is most likely to represent the actual gate behavior, thereby allowing the gate to be tested and identified. Subsequently, the system transitions to another desired gate configuration by loading a new layer set, which is $H_3$ here. To confirm the realization of the intended gate, another round of process tomography and reconstruction are performed. The annotations at the initiation and termination points of the tomography arrow in Fig. 5a signify the initial and measured tomography matrices, respectively. Meanwhile, the annotations along the switch arrow denote the previous and updated layers.

After configuring to the desired gate, the next step involves optimizing the spacing between each layer to match the experimental setup. Adjusting the spacing between SLM2 and the mirror can be a tedious task, but it can be avoided by varying the spacing during pattern generation. This approach is more compatible with automatic protocols. First, a rough estimate of the spacing range is made based on the setup, and then five equally spaced sampling points are selected as the initial condition. The visibilities of these sampling points and their distribution are used to update the sampling points for the next step (see Supplementary Note 5). This iterative process stops when the searching range is below the threshold, where the spacing variance does not correspond to a significant change in visibility. The threshold is presumed to be lower than 0.1 mm, which is supported by the analysis of Z-axis offset effects in Supplementary Note 6 (with some approaches to improve misalignment tolerance). Fig. 5c shows the visibilities of different sampling spots in each update, with the highest visibility observed around 0.97 at 41 mm.

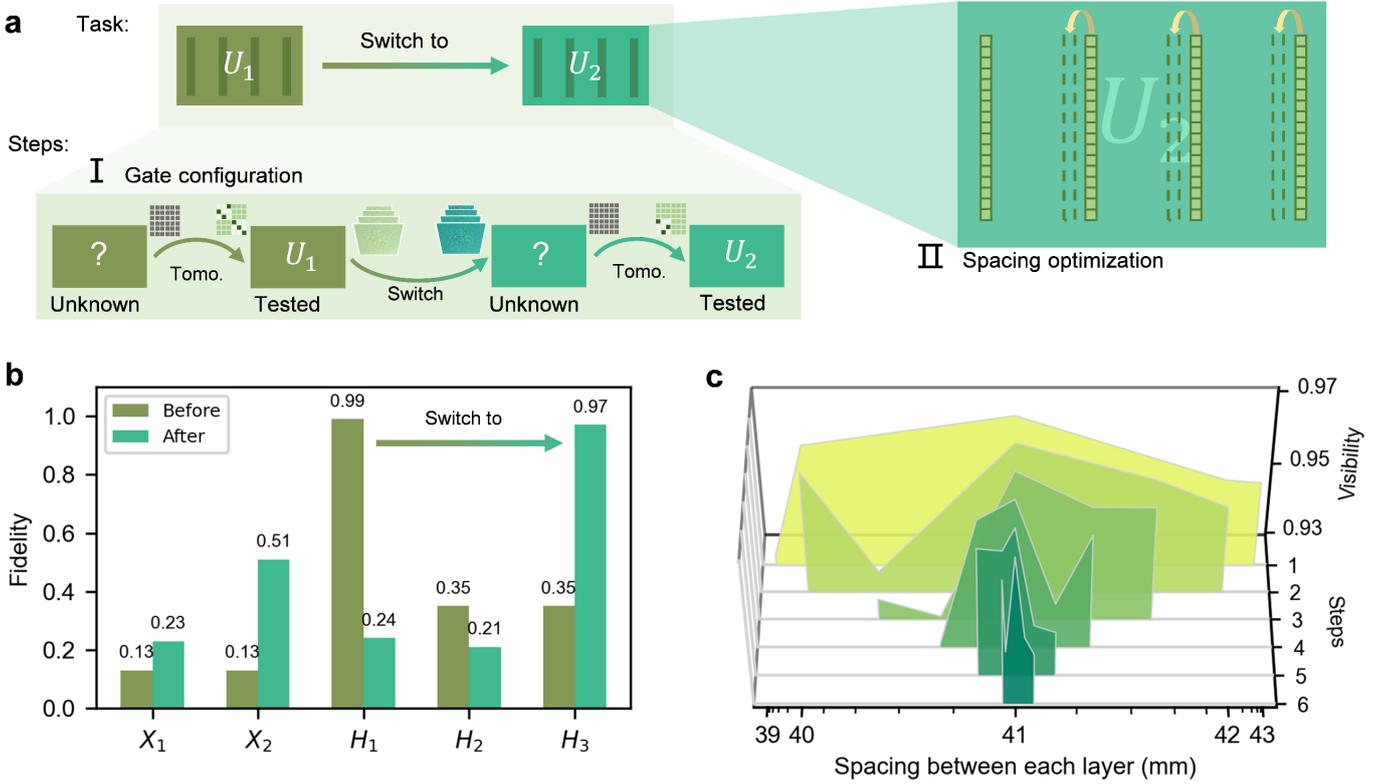

**Fig. 5 Intelligent deployment of the reconfigurable spatial mode quantum gate. a** Schematic of the self-configuration and optimization process. Tomo.: tomography; The process involves testing the current gate via process tomography, switching to the desired gate by loading a new layer set. To confirm the realization of the intended gate, another round of process tomography is performed. The annotations at the initiation and termination points of the tomography arrow signify the initial and measured tomography matrices, respectively. The annotations along the switch arrow denote the previous and updated layers. Once the configuration is complete, a spacing optimization process is performed to match the calculation spacing with the physical spacing. **b** Gate fidelities used to test the gate in the self-configuration demonstration before ($H_1$ gate) and after switching ($H_3$ gate). **c** Visibility plot of the search steps for the optimal spacing between each layer. A rough range is estimated first, followed by a search process to narrow down the possible interval of maximum visibility. The maximum visibility achieved is ~0.97 at 41 mm.

**Gate performance analysis and comparison with WFM**

Optimizing practical parameters inherent to the SLM is of paramount importance to achieve optimal quantum gate performance. To assess the critical pixel number of the SLM, it is essential to explore two distinct scenarios, as diagramed in Fig. 6a-b. The first scenario involves variations in the number of pixels while keeping the physical dimensions of the layers fixed. Alternatively, the second scenario considers changes in the number of pixels while maintaining the pixel dimensions, which, in turn, alter the physical dimensions of the layers.

In the first scenario, our primary focus is on the influence of pixel density or sampling precision on quantum gates. Our analysis intuitively suggests that higher pixel densities contribute to enhanced quantum gate performance. Additionally, we delve into the consequences of upscaling or downsampling quantum gates with different pixel densities. The findings depicted in Fig. 6e reveal that both upscaling and downsampling have a detrimental effect on performance, although they operate through distinct mechanisms. Upscaling reduces performance by diminishing the effective modulation degrees of freedom, while downsampling leads to performance degradation due to sampling errors arising from pixel merging. In the second scenario, our analysis demonstrates, in Fig. 6f, that increasing the number of pixels can indeed boost performance. However, it's important to note that the marginal improvement diminishes, and performance changes become less significant after reaching approximately 384 pixels. This suggests that, for the scope of our work, selecting 384 pixels suffices to achieve the desired outcomes.

Moreover, the gray level depth of the SLM can significantly impact quantum gate performance, as illustrated in Fig. 6c and g. Lower gray level depths introduce inaccuracies in pixel values. Surprisingly, our findings indicate that quantum gate performance remains relatively stable within the range of 32 to 512 gray levels, with a noticeable performance drop occurring only when the depth falls below 16 levels. Another critical parameter affecting performance is the presence of phase distortion in the SLM, as shown in Fig. 6d and h. The distortion phase applied is a combination of the fourth and fifteenth terms of Zernike polynomials. Phase distortion introduces additional phase components that render the original design ineffective. Consequently, the necessity of pre-compensating for shape distortions in the SLM becomes evident.

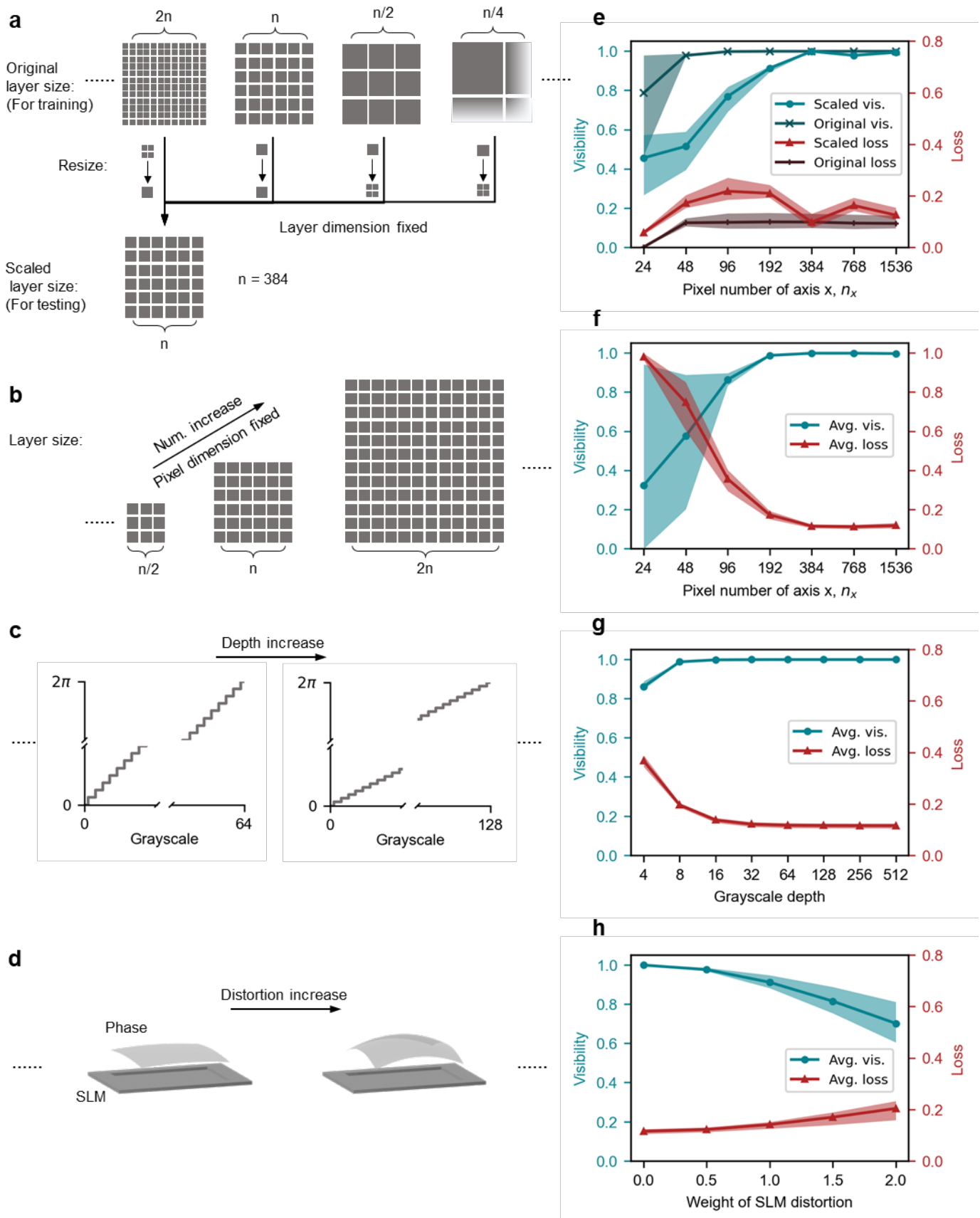

**Fig. 6** Gate performance across SLM parameters. The diagrams of **a** pixel number scaling with layer dimensions fixed, **b** pixel number scaling with pixel dimensions fixed, **c** grayscale and **d** distortion of SLM variations. **e-h** Corresponding visibility and loss plots corresponding to the parameters explored in **a-d**. Both visibility and loss metrics demonstrate degradation when either upscaling or downsampling is applied. An increase in pixel numbers enhances performance, but with diminishing marginal returns, validating our choice of 384 pixels as sufficient. Notably, a decrease in grayscale depth exhibits negligible effects on performance until it falls below 16 levels. Furthermore, SLM distortion introduces undesired phase shifts to the D$^2$NN pattern, resulting in reduced performance.

Despite our successful experimental demonstration of the D$^2$NN spatial mode quantum gate, we remain concern about its upper performance limit. Two primary metrics are used to evaluate the device performance: visibility and loss. Visibility characterizes the quality of the output, and is defined as $V_i = |\langle \varphi_i | \varphi_i \rangle|^2 / \sum_{i,j}^{d} |\langle \varphi_i | \varphi_j \rangle|^2$, where $|\varphi_j\rangle$ is the defined computational basis, and $d$ is the dimension. Loss, defined as $(|E_{in}|^2 - |E_{out}|^2)/|E_{in}|^2$, is the normalized energy waste and negatively related to efficiency. It is worth mentioning that the energy loss here referring to the scattering loss that arises from undesired phase pattern design or pattern misalignment, and does not account for Fresnel reflection effects. Several factors might impact visibility and loss, including the number of trainings, the spacing between each layer, and the number of layers with a fixed spacing or a fixed total length, as presented in Fig. 7a-d. Our data indicate that performance generally improves with increasing numbers of training and layers, eventually converging to a certain value. It is intuitive that the D$^2$NN fitness shows positive correlation with the training number, resulting in better visibility and lower loss. Similarly, more layers provide more DoFs to achieve desired conversion, whether fixed spacing or fixed total length. It is worth noting that poor and less reliable performance is observed when the number of trainings or layers is very small, as these designs do not fully function. Specifically, some inputs are not correctly converted, and their energy is lost due to scattering. Of interest, the loss related to spacing between each layer (Fig. 7b) shows a small dip around 10 mm and flattens out as the spacing increases. This discrepancy could be attributed to lower pixel utilization for shorter diffraction distances and larger losses due to scattering for larger diffraction distances.

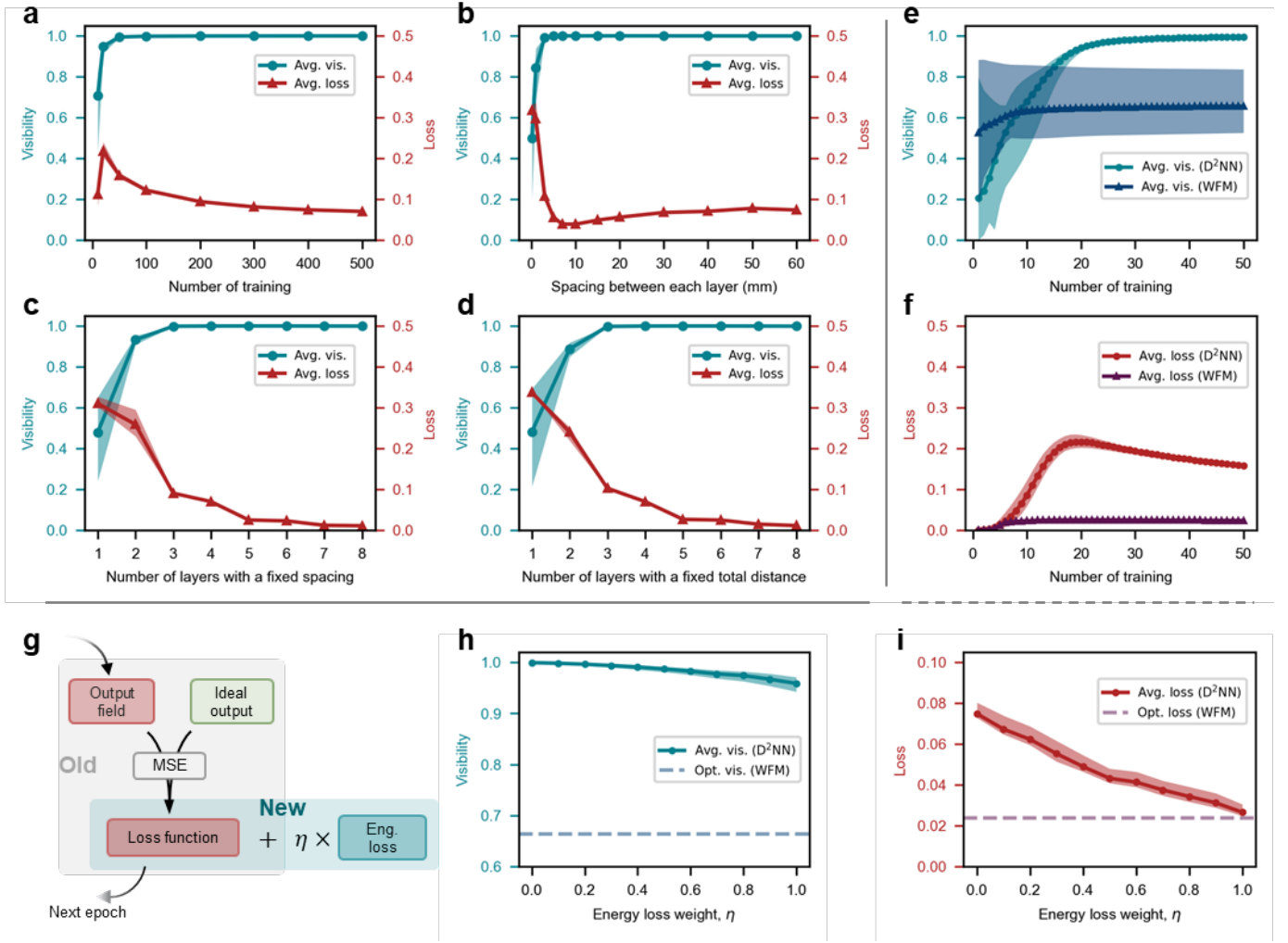

**Fig. 7 Performance evaluation of three-dimensional $X_1$ quantum gate.** Simulated visibility (blue) and loss (red) as a function of **a** the number of trainings (4 layers)**, b** spacing between each layer (4 layers), **c** number of layers with a fixed spacing, and **d** number of layers with a fixed total length. **b-d** are trained for 500 epochs. **e-f** Performance comparison of $D^2NN$ and WFM in visibility and loss. **g** Introduction of a separate energy loss term to emphasize its expression during the training process. **h** Visibility and **i** loss metrics based on varying weights assigned to the energy loss component within the loss function. All 12 states in three-dimensional MUBs are considered. The solid curves represent the average data, and the filled areas show the range of values for different states.

To clearly illustrate the improvement of our approach, we compare the performance of $D^2NN$ with WFM in Fig. 7e-f. The results indicate that both methods converge within a few dozen iterations in terms of visibility and loss. WFM achieves near-perfect energy conservation, as evidenced by its impressive loss plot. This can be explained by the weight update mechanism in WFM, which is based on the overlap between the forward propagation field and the backpropagation field. In contrast, $D^2NN$ allows for arbitrary connections between nodes and uses the Adam algorithm to avoid local optima[46]. Although $D^2NN$ requires more training to reduce its loss, it outperforms WFM in terms of visibility and approaches the ideal value of 1. It is worth noting that in all cases shown in Fig. 7, we use all 12 states in the three-dimensional MUBs as input states, which diminishes the performance of WFM. Overall, the comparison reveals that D$^2$NN significantly improves visibility at a small cost of energy loss.

Fortunately, theoretical energy losses associated with D$^2$NN can be controlled by incorporating them into the optimization loss function, as depicted in Fig. 7g. Originally, the loss function is defined as the MSE between the inferred outputs $E$ and the corresponding theoretical outputs $\hat{E}$, which is $f_L = \frac{1}{n^2} \sum_{i,j}^{n} \left( \left| E_{i,j} - \hat{E}_{i,j} \right|^2 \right)$. It is worth noting that the original loss function inherently accounts for energy loss to some extent, yet the energy term constraints are not fully expressed during the training process. Instead, greater emphasis is placed on optimizing visibility. The introduction of a separate energy loss term aims to strike a balance of these two parameters. As the weight assigned to the energy loss in the new loss function increases, we observe a slight reduction in output visibility. However, it is noteworthy that the D$^2$NN outputs continue to outperform the optimized visibility achieved with the WFM method. Simultaneously, the energy loss of D$^2$NN converges towards that of the WFM method, as shown in Fig. 7h and i, underscoring the superiority of our approach.

## Discussion

In conclusion, we present a universal approach utilizing deep learning to design high fidelity and high-dimensional spatial mode quantum gates. Our scheme is implemented with a group of D$^2$NN-generated phase layers, allowing us to experimentally demonstrate all three-dimensional quantum gates and a single-photon *CNOT* gate with an average fidelity over 99%, as characterized by quantum process tomography. Compared to discrete optics demonstrations, our implementation is more practical, relatively compact, and capable of integrating multiple operations into a single device, simplifying the physical implementation of quantum circuits and reducing error rates. Moreover, our approach enables the cascading of multiple gates, facilitating the implementation of quantum circuits for other algorithms. Our design is also reconfigurable and allows for closed-loop control, enabling automatic configuration[47]. Finally, we analyze the theoretical performance upper limit of D$^2$NN gates and compare it with the WFM approach. Although we find some similarities between the two, D$^2$NNs are more general and can achieve higher visibility by varying the loss factor.

The experimental performance of the gates proposed in this work is affected by the gray-level accuracy of the phase modulation device and its diffraction efficiency, both of which are expected to be improved with advancements in device manufacture. Misalignment of the experimental setup is another factor that may lead to deviations from the ideal output states and energy losses. This issue could be addressed through better alignment protocols, auto-alignment algorithms, or self-optimization processes. Other approaches to reduce the impact of misalignment include modeling the entire physical process more precisely and implementing pre-compensation techniques, or introducing random offsets in the D$^2$NN design to improve tolerance[48,49] (see Supplementary Note 7 for further details). Regarding the measured experimental loss in the SLM, accounting for the four reflections, the observed 63.44% loss aligns reasonably well with the SLM's inherent efficiency of approximately 90%. To address this concern in reflective setups, a potential solution involves replacing the SLM with manufactured phase layers coated with high reflection materials[50]. For configurations with a transmissive nature, it is worth considering potential collaboration between certain anti-reflection techniques and the D$^2$NN architecture [51]. Moreover, the theoretical upper limit of D$^2$NN constrains the gate performance, but it can be improved by using different D$^2$NN architectures[52,53], various layers (e.g., nonlinear layers[54,55], transmissivity modulation layers[32,56]), operations in Fourier space[54], or special training strategies[57].

Regarding scalability, the spatial-mode-encoding scheme enables access to a high-dimensional Hilbert space in theory[58]. Previous experiments have achieved dozens or even hundreds of dimensions in the coding space using similar multi-plane devices[40,59,60]. It has also been shown that the dimensionality is linearly proportional to the number of phase planes[61]. These findings suggest the potential of our spatial mode gate implementation to handle qudits in higher dimensions at the cost of increasing the number of layers. Furthermore, we test several coding bases for the three-dimensional gates and the *CNOT* gate, and they all exhibit reliable behavior, indicating that our approach has excellent robustness to different spatial modes of encoding, even with a certain number of dimensions.

## Materials and methods

**Experimental setup**

A 1550 nm laser (~500 mW) and the PPLN1 are used to generate the pump beam at 775 nm (~2 mW) for the SPDC process in PPLN2, as shown in the left bottom of Fig. 1b. A set of bandpass filters are used after the PPLN1 to suppress the residual 1550 nm laser. The 775 nm pump beam is focused by a lens (f=150 mm, not shown) at the PPLN2 with a radius of 70 μm to enhance SPDC efficiency. To filter out the 775 nm pump beam, two 1550 nm bandpass filters, each with a 10 nm bandwidth, are utilized. Both crystals satisfy type-0 phase matching condition and are mounted on the temperature-controlled stages to uphold the optimal phase matching temperature: 56.6 °C for PPLN1 and 82 °C for PPLN2. Accounting for system losses, detector efficiency, and a heralding efficiency of approximately 14.5%, the characterized source brightness stands at ~ $3.79 \times 10^6$ pairs/s/nm/mW. Photon pairs generated via SPDC are separated by a BS, and then one of them is coupled into free-space by a collimator (Thorlabs F260FC-1550) at a beam waist of 1.5 mm. A fiber polarization controller (FPC1) is used to align the photon polarization to the working direction of an SLM (PLUTO-2.1-TELCO-013, SLM1 in Fig. 1b) which carves out required LG modes by imprinting complex modulation phase patterns. A 4-f lens set (L1 and L2) with a pinhole is required by the complex modulation method to filter out the desired diffraction light. A mirror (Thorlabs MRA10-M01) is set opposing the SLM2 so that light bounces off SLM2 four times to transform input modes under the configuration of the generated phase layers on SLM2. The four modulation phase layers are 384×384 pixels in size (8 μm) and spaced 41 mm apart. Owning to its reflection implementation and tilted incident light axis, the spacing between SLM2 and mirror is even less than half of 41 mm, making its footprint relatively compact. At the output plane, another 4-f lens set (L3 and L4) is used to image the outputs onto the SLM3, which, together with the third 4-f imaging system, a collimator, a single-mode fiber, and the superconducting nanowire single-photon detector (Eos4 of Single Quantum, 70% efficiency at 1550 nm, ~300 dark counts), constitute a spatial mode analyzer. A multichannel picosecond event timer (HydraHarp 400 of PicoQuant) is used to record all detector clicks for later analysis.

## Acknowledgements

The authors would like to extend sincere thanks to Dr. Carmelo Rosales-Guzmán and Dr. Xiang Cheng for their invaluable insights and thoughtful discussion during the development of this work. This work was supported by the National Natural Science Foundation of China (62125503, 62001182, 62261160388, 62371202), the Natural Science Foundation of Hubei Province of China (2023AFA028, 2023AFB814), the Key R&D Program of Guangdong Province (2018B030325002), the Key R&D Program of Hubei Province of China (2021BAA024, 2020BAB001), the Shenzhen Science and Technology Program (JCYJ20200109114018750), and the Innovation Project of Optics Valley Laboratory (OVL2021BG004).

## Conflict of interests

There are no conflicts of interest in this work.

# Contributions

J.W. and Q.W. developed the concept and conceived the experiments. Q.W. and J.L. designed the optical system. Q.W. and D.L. simulated the system and carried out the experiment with J.L. Q.W., D.L., and J.L. prepared the experimental data analyses. Q.W. drafted the manuscript with support by all co-authors. J.W. finalized the paper. J.W. supervised the project.

Supplementary Information for

Ultrahigh-Fidelity Spatial Mode Quantum Gates in High-Dimensional Space by Deep Diffractive Neural Networks

Qianke Wang[1,2,†], Jun Liu[1,2,†], Dawei Lyu[1,2], Jian Wang[1,2,*]

[1] *Wuhan National Laboratory for Optoelectronics and School of Optical and Electronic Information, Huazhong University of Science and Technology, Wuhan 430074, Hubei, China*

[2] *Optics Valley Laboratory, Wuhan 430074, Hubei, China*

[†] *These authors contributed equally to this work.*

*\* Corresponding Author:* [jwang@hust.edu.cn](mailto:jwang@hust.edu.cn)

**Supplementary Note 1: Phase pattern generation**

We employ the TensorFlow 2.5.0 architecture to model the physical diffraction layers and train the network models. Each layer is defined as a complex value matrix, which is multiplied by the incoming field as the modulation process. The angular spectrum method is used to describe the free-space propagation, and the mean squared error between the ideal output field and the inference is defined as the loss function. The $D^2NN$ architecture is shown in Fig. S1. To compensate for systematic errors, we introduce random offsets and modulation blurring correction to the complex-valued layer model, which are discussed in detail in Supplementary Note 6 and Note 7. We use the Adam optimizer with default parameters in TensorFlow, except that the learning rate was set to 0.01. Adam combines the advantages of adaptive learning rates from RMSProp and the momentum from SGDM. It adapts the learning rates per parameter and includes a moving average of past gradients. Adam also adds bias-correction compared to RMSProp. The comparison of the accuracy versus training numbers of different optimizers are plotted in Fig. S2. It should be noted that this learning rate may not be optimal, and further exploration could improve network performance.

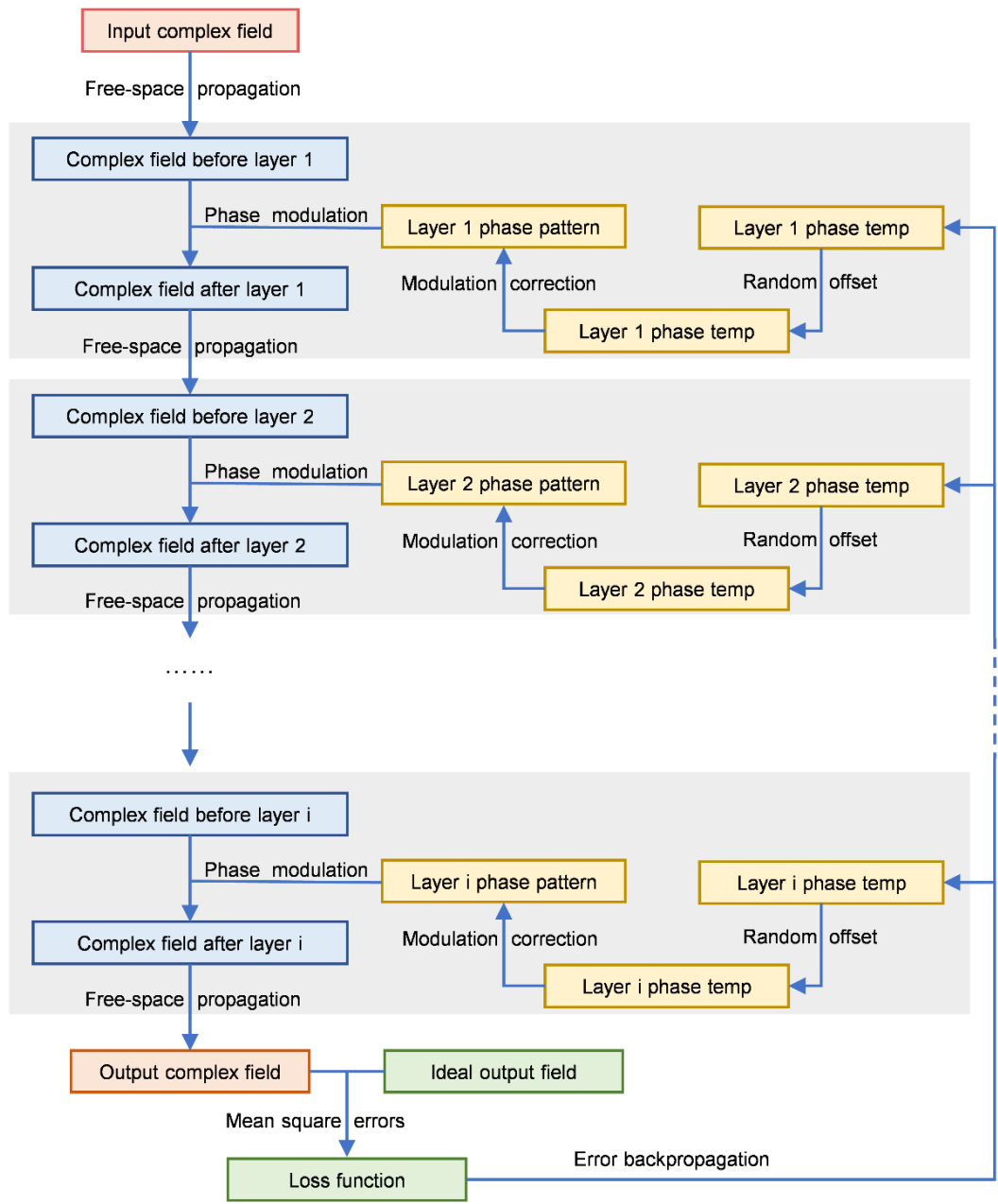

**Fig. S1.** Architecture of D²NN using TensorFlow.

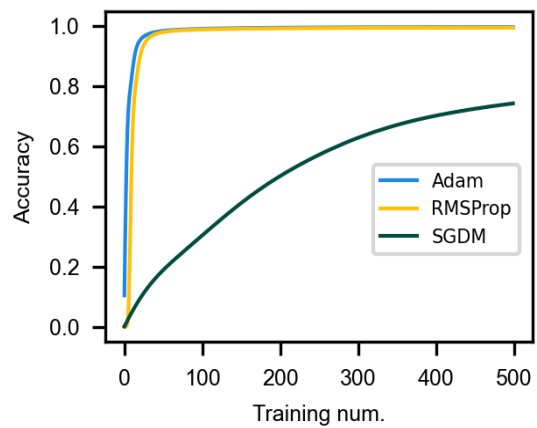

**Fig. S2.** Accuracy versus training number of SGDM (learning rate= 0.1, momentum=0.9), RMSProp (learning rate=

**Supplementary Note 2: Matrix representation of quantum gates, MUBs and complement results**

In the three-dimensional space, the Pauli X gate is no longer limited to flipping two computational bases, but rather cyclic shift-transformation of three bases as the matrix representation described in Eq. (S1). Operation $X_0$ is the identity matrix and shifts no input. $X_1$ and $X_2$ shift the input basis states once and twice, respectively.

$$X_0 = \begin{pmatrix} 1 & 0 & 0 \\ 0 & 1 & 0 \\ 0 & 0 & 1 \end{pmatrix}, X_1 = \begin{pmatrix} 0 & 1 & 0 \\ 0 & 0 & 1 \\ 1 & 0 & 0 \end{pmatrix}, X_2 = \begin{pmatrix} 0 & 0 & 1 \\ 1 & 0 & 0 \\ 0 & 1 & 0 \end{pmatrix} \quad (S1)$$

The matrix representation of three-dimensional Hadamard gates is described by eq. (S2), where $\omega = e^{2\pi i/3}$.

$$H_0 = \begin{pmatrix} 1 & 0 & 0 \\ 0 & 1 & 0 \\ 0 & 0 & 1 \end{pmatrix}, H_1 = \frac{1}{\sqrt{3}}\begin{pmatrix} 1 & 1 & 1 \\ 1 & \omega & \omega^2 \\ 1 & \omega^2 & \omega \end{pmatrix}, H_2 = \frac{1}{\sqrt{3}}\begin{pmatrix} 1 & 1 & 1 \\ \omega & \omega^2 & 1 \\ \omega & 1 & \omega^2 \end{pmatrix}, H_3 = \frac{1}{\sqrt{3}}\begin{pmatrix} 1 & 1 & 1 \\ \omega^2 & 1 & \omega \\ \omega^2 & \omega & 1 \end{pmatrix} \quad (S2)$$

According to ref. 1, all inequivalent sets of MUBs in three dimensions could be derived from the Hadamard matrices. The state vectors $|\psi_i\rangle$ are the columns of these Hadamard matrices, thus 12 states in total. It is convenient to verify by calculating $|\langle \psi_i | \psi_j \rangle|^2 = 1/d$, where $d$ is the dimension.

The *CNOT* gate operating on 2×2-dimensional space has more complicated MUBs. We construct an over-complete state set from the column vectors in matrix $S$ which is the tensor product of six two-dimensional MUBs, as shown in eq. (S3). This state set with 36 state vectors oversteps the range of four-dimensional MUBs but still works in quantum process tomography.

$$C = \begin{pmatrix} 1 & 0 & \frac{1}{\sqrt{2}} & \frac{1}{\sqrt{2}} & \frac{1}{\sqrt{2}} & \frac{1}{\sqrt{2}} \\ 0 & 1 & \frac{1}{\sqrt{2}} & \frac{-1}{\sqrt{2}} & \frac{i}{\sqrt{2}} & \frac{-i}{\sqrt{2}} \end{pmatrix} \quad (S3)$$

$$S = C \otimes C$$

As a complement to the results presented in the main text, we characterize the rest of the three-dimensional gates, including the $X_2$, $H_2$, and $H_3$ gates. Fig. S3a-c presents the tomography matrices of these three gates, and Fig. S3d-f are the corresponding reconstructed process matrices $\chi$. The fidelities are around 97% revealing decent performance.

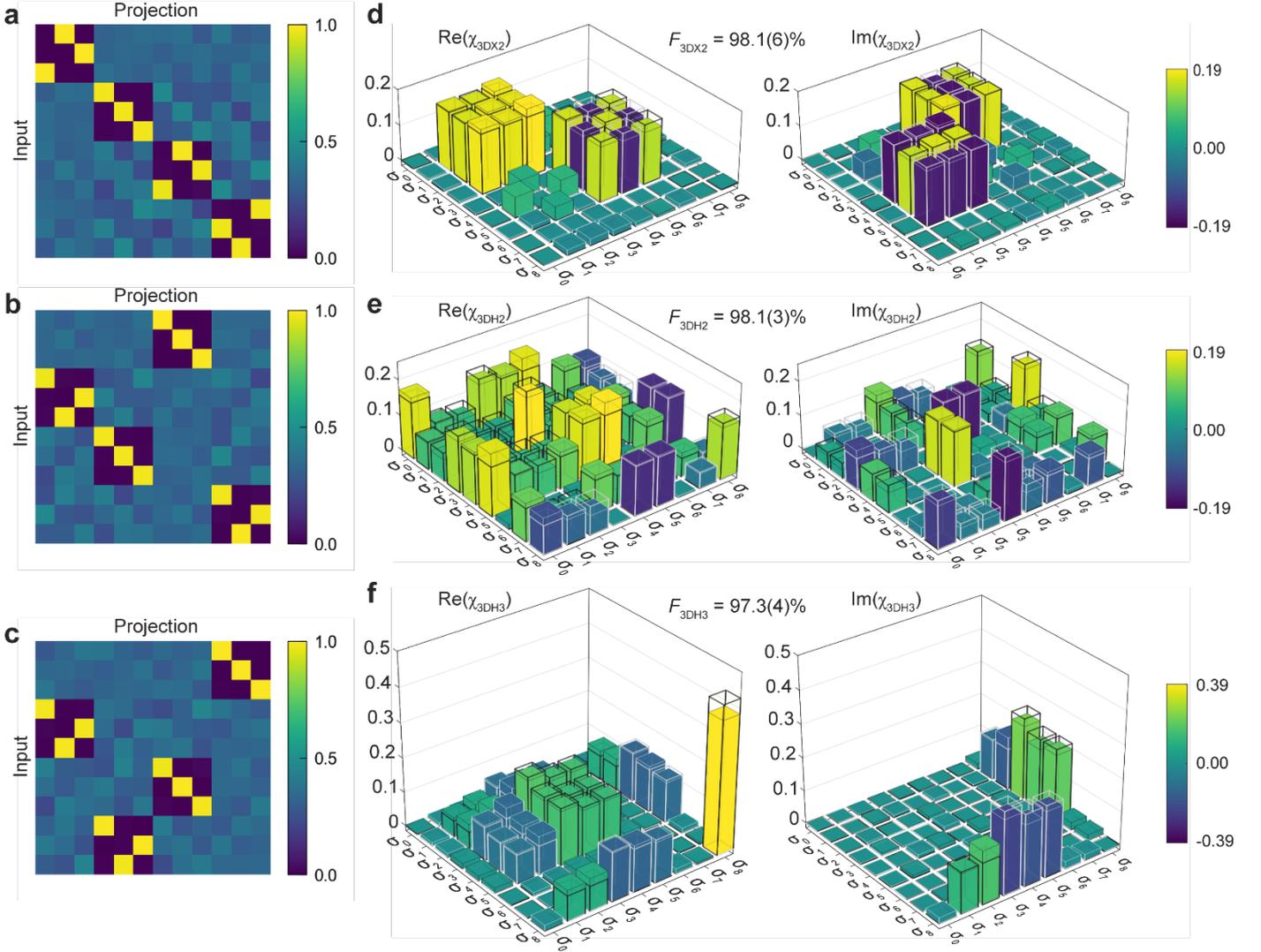

**Fig. S3. Quantum process tomography of spatial mode quantum gates. a-c** Tomography matrices and **d-f** reconstructed process matrices $\chi$ for the $X_2$ gate, $H_2$ gate and $H_3$ gate.

## Supplementary Note 3: Quantum process tomography reconstruction with maximum-likelihood-estimation method

The maximum-likelihood-estimation principle as a mature method has been applied in quantum state tomography and quantum process tomography for a long time. Here we briefly introduce the procedure of it. Considering an unknown quantum process $E$ is a linear completely positive map $M_E$ from the Hilbert space $H$ to the Hilbert space $K$, we can test it with a bunch of probe states $\rho_m$, and the output states $\rho_{out} = M_E(\rho_m) = \text{Tr}_H(E \rho_m^T \otimes I_K)$, where $\text{Tr}_H$ is the partial trace of space $H$, and $I_K$ is the identity operator on space $K$, and $T$ denotes the transposition. The trace-preserve (TP) condition leads to $\text{Tr}_K(\rho_{out}) = \text{Tr}_H(\rho_m)$ for any $\rho_m$, and the operator $E$ must satisfy

$$\mathrm{Tr}_K(E) = I_H \tag{S4}$$

where $I_H$ is the identity operator on space $H$. Measurements $\Pi_n$ are carried out on the corresponding output states, yielding theoretical probabilities $p_{mn} = \mathrm{Tr}(\rho_{out}\Pi_n) = \mathrm{Tr}(E\rho_m^T \otimes \Pi_n)$. Let $f_{mn}$ denote the experimental frequency of detecting $\rho_m$ with $\Pi_n$, and the estimated operator $E$ should maximize the log-likelihood function

$$L(f_{mn}, p_{mn}(E)) = \sum_{m,n} f_{mn} \ln p_{mn} - \mathrm{Tr}(\Lambda E) \tag{S5}$$

where $\Lambda = \lambda \otimes I_K$, and $\lambda$ is the Lagrange multiplier in matrix form to satisfy the TP condition of Eq. (S4). Varying Eq. (S5) with respect to $E$ will give the extremal equation for operator $E$

$$\begin{aligned} L(f_{mn}, p_{mn}(E+\delta E)) - L(f_{mn}, p_{mn}(E)) &= 0 \\ \mathrm{Tr}\left(\left(\sum_{m,n} \frac{f_{mn}}{p_{mn}}\rho_m^T \otimes \Pi_n - \Lambda\right)\delta E\right) &= 0 \end{aligned} \tag{S6}$$

Eq. (S6) holds for all $\delta E$, then

$$\begin{aligned} R &= \Lambda, \quad (R = \sum_{m,n} \frac{f_{mn}}{p_{mn}}\rho_m^T \otimes \Pi_n) \\ \Lambda^{-1}RE &= E, \text{ or} \\ ER\Lambda^{-1} &= E \end{aligned} \tag{S7}$$

Combining two variations of Eq. (S7), the symmetrical expression suitable for iterations is derived

$$E_{i+1} = \Lambda_i^{-1} R_i E_i R_i \Lambda_i^{-1}, \quad \lambda_i = \left(\mathrm{Tr}_K(R_i E_i R_i)\right)^{\frac{1}{2}} \tag{S8}$$

where $i$ denotes the $i$-th iteration. This expression preserves the positive semi-definiteness and trace normalization of the operator since $R_i E_i R_i$ is positive-definite and the iterations satisfy $\mathrm{Tr}_K(E) = I_H$. Let $E_0 = I_{H\otimes K}/d_K$, where $d_K$ is the dimension of space $K$, the estimated operator $E_i$ will gradually approach the theoretical operator of the test results during the numerical iterations.

**Supplementary Note 4: Example of the Deutsche algorithm**

As can be seen from Fig. S4, the Deutsche algorithm consists of four steps:

1. Prepare input states. The first qubit labeled as $x$ is initialized to $|0\rangle_x$, and the second qubit $y$ is initialized to $|0\rangle_y$. Then apply $H$ gates to each qubit, yielding

$$|\psi_1\rangle = \frac{|0\rangle_x + |1\rangle_x}{\sqrt{2}} \otimes \frac{|0\rangle_y - |1\rangle_y}{\sqrt{2}} \tag{S9}$$

2. Apply the quantum oracle. The oracle maps the input state $|x\rangle|y\rangle$ to $|x\rangle|y \oplus f(x)\rangle$, thus the output is

$$|\psi_2\rangle = \frac{|0\rangle_x + |1\rangle_x}{\sqrt{2}} \otimes \frac{|0\rangle_y \oplus f(x) - |1\rangle_y \oplus f(x)}{\sqrt{2}}$$
$$= (-1)^{f(x)} \frac{|0\rangle_x + |1\rangle_x}{\sqrt{2}} \otimes \frac{|0\rangle_y - |1\rangle_y}{\sqrt{2}} \tag{S10}$$

When the qubit $x$ is $|0\rangle_x$, the corresponding $f(x)$ is $f(0)$, and vice versa. Note the $x$ and $y$ used in the Dirac notation only indicate the qubits and do not have any assigned value, whereas the $x$ and $y$ used as inputs in the oracle have assigned values. So, the Eq. (S10) could be

$$|\psi_2\rangle = \frac{(-1)^{f(0)}|0\rangle_x + (-1)^{f(1)}|1\rangle_x}{\sqrt{2}} \otimes \frac{|0\rangle_y - |1\rangle_y}{\sqrt{2}} \tag{S11}$$

3. Apply an $H$ gate to each qubit. The final state will be

$$|\psi_3\rangle = |f(0) \oplus f(1)\rangle_x \otimes |1\rangle_y = \begin{cases} |0\rangle_x \otimes |1\rangle_y, & f(x) \text{ is constant} \\ |1\rangle_x \otimes |1\rangle_y, & f(x) \text{ is balanced} \end{cases} \tag{S12}$$

4. Measure the final state $|\psi_3\rangle$. Theoretically, only the measurement of qubit $x$ is necessary, and a constant function will be measured as $|0\rangle_x$ and a balanced function as $|1\rangle_x$.

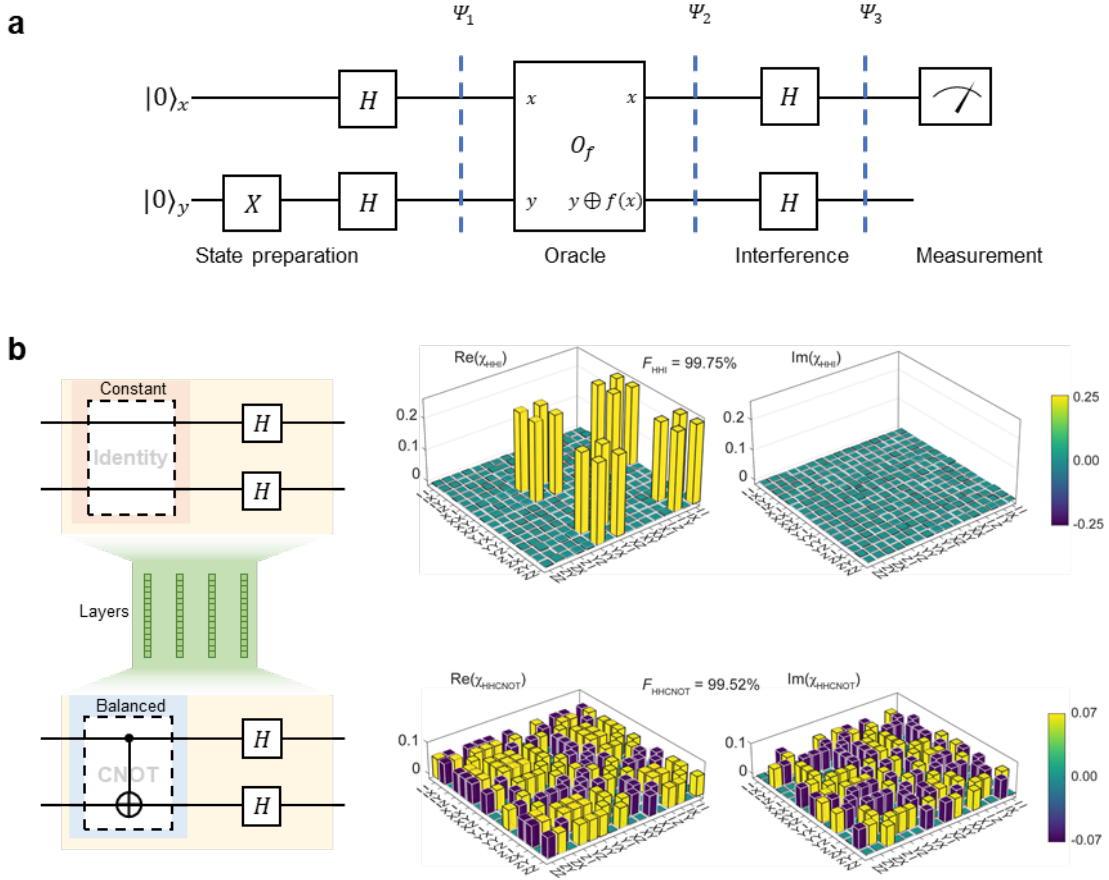

**Fig. S4. Deutsch algorithm diagrams. a** Quantum circuit for the Deutsch algorithm. The circuit can be subdivided into four main components. The oracle function denoted as $O_f$ applies identity operator to qubit $x$, while applying $XOR$ with $f(x)$ to qubit $y$. **b** Combinations of the Oracle operations and interference operations in a single D²NN and corresponding process matrices.

The *CNOT* gate is an example that implements a balanced function:

$$\begin{aligned}
CNOT\left(\frac{|0\rangle_x+|1\rangle_x}{\sqrt{2}} \otimes \frac{|0\rangle_y-|1\rangle_y}{\sqrt{2}}\right) &= CNOT\frac{1}{2}(|0\rangle_x+|1\rangle_x)(|0\rangle_y-|1\rangle_y) \\
&= \frac{1}{2}|0\rangle_x(|0\rangle_y \oplus f(0)-|1\rangle_y \oplus f(0)) \\
&\quad + \frac{1}{2}|1\rangle_x(|0\rangle_y \oplus f(1)-|1\rangle_y \oplus f(1)) \\
&= \frac{1}{2}(|0\rangle_x-|1\rangle_x)(|0\rangle_y-|1\rangle_y) \\
\xrightarrow{H \otimes H} &= \frac{1}{2}|1\rangle_x|1\rangle_y
\end{aligned} \quad (S13)$$

Similarly, the identity operation is an example of a constant function implementation:

$$\begin{aligned}
I\left(\frac{|0\rangle_x+|1\rangle_x}{\sqrt{2}} \otimes \frac{|0\rangle_y-|1\rangle_y}{\sqrt{2}}\right) &= I\frac{1}{2}(|0\rangle_x+|1\rangle_x)(|0\rangle_y-|1\rangle_y) \\
&= \frac{1}{2}(|0\rangle_x+|1\rangle_x)(|0\rangle_y-|1\rangle_y) \\
\xrightarrow{H \otimes H} &= \frac{1}{2}|0\rangle_x|1\rangle_y
\end{aligned} \quad (S14)$$

The characteristic of implementing a simple combination of multiple basic quantum gates in a single D²NN is also showcased through the application of the Deutsch algorithm. In essence, the D²NN functions by fitting the transformation matrix of the inputs and outputs. For a given number of dimensions, such as the case with four dimensions here, a combination of multiple basic quantum gates still results in a four-dimensional unitary matrix, thereby maintaining the same fitting complexity for the D²NN. To provide specific details, the gate loaded onto SLM2 is represented as $(H \otimes H)I_4$ for the constant situation and $(H \otimes H)CNOT$ for the balanced situation. Here, $H$ signifies the two-dimensional Hadamard gate, and $I_4$ represents the four-dimensional identity matrix. The process matrices for these two situations are visually presented in Fig. S4b. Additionally, the compression of a more extensive two-qubit circuit or higher dimensions, are also possible.

**Supplementary Note 5: Update rules for spacing optimization**

Finding the optimal spacing that maximizes performance is an optimization problem that is difficult to solve for physical systems. Fortunately, in our case, finding a local optimum is sufficient. To achieve this, we use a search protocol that scans a given range with a certain precision. The pseudo-code for the protocol is as follows.

**Algorithm 1:** Our update rules for spacing optimization. Function $\text{idx}(x)$ gives the index of $x$ in an array. Function $V(s)$ gives the output state visibility of the spacing $s$

**Require:** $\{R_{e,n}\}$, $n \in \{0,1\}$, $R_{e,0} < R_{e,1}$: Estimated searching range

**Require:** $\{R_{T,n}\}$, $n \in \{0,1\}$, $R_{T,0} < R_{T,1}$: Searching range threshold

**Require:** $s_T$: Spacing threshold

1:     $\{R_{0,n}\} \leftarrow \{R_{e,n}\}$ (Initialize searching range)

2:     $i \leftarrow 0$ (Initialize step number)

3:     **while** $R_{T,0} < s_{i,m} < R_{T,1}$ && $s > s_T$, $(m \in \{0,1,2,3,4\})$

4:        $s_{i,m} \leftarrow R_{i,0} + m \cdot (R_{i,1} - R_{i,0})/4$

5:        **if** $\text{idx}(\min(V(s_{i,m}))) == 0$ && $\text{idx}(\max(V(s_{i,m}))) == 4$

6:           $R_{i,0} \leftarrow s_{i,0}$

7:           $R_{i,1} \leftarrow s_{i,0} + 2 \cdot (s_{i,4} - s_{i,0})$

8:        **else if** $\text{idx}(\min(V(s_{i,m}))) == 4$ && $\text{idx}(\max(V(s_{i,m}))) == 0$

9:           $R_{i,0} \leftarrow s_{i,4} - 2 \cdot (s_{i,0} - s_{i,4})$

10:          $R_{i,1} \leftarrow s_{i,4}$

11:       **else if** $0 < \text{idx}(\max(V(s_{i,m}))) < 4$

12:          $d \leftarrow \text{idx}(\max(V(s_{i,m})))$

13:          $R_{i,0} \leftarrow s_{i,d-1}$

14:          $R_{i,1} \leftarrow s_{i,d+1}$

15:     **end if**

16:     $i \leftarrow i+1$

17: **end while**

## Supplementary Note 6: Misalignment analysis

Misalignment is an inevitable factor that affects the experiment. To mitigate its impact, it is important to understand different types of misalignment. In our experiment, we observe offsets that usually arise in the direction perpendicular (Z-axis) and parallel (X-axis) to the layers, as shown in the upper left legend of Fig. S5a. The Y-axis offset behavior is expected to be similar to the X-axis and is therefore neglected. To evaluate the degradation caused by misalignment, we plot the simulated visibility and energy loss of the output states and plot the field profiles for better visualization. Fig. S5a-b show intuitive results and lager offsets contribute to worse performance. It can be seen from this figure that the tolerances to both kinds of offsets are very small. To enhance the robustness of the system, we introduce a spacing optimization process (described in the main text) to address Z-axis misalignment, and a random offset method to address X-axis misalignment.

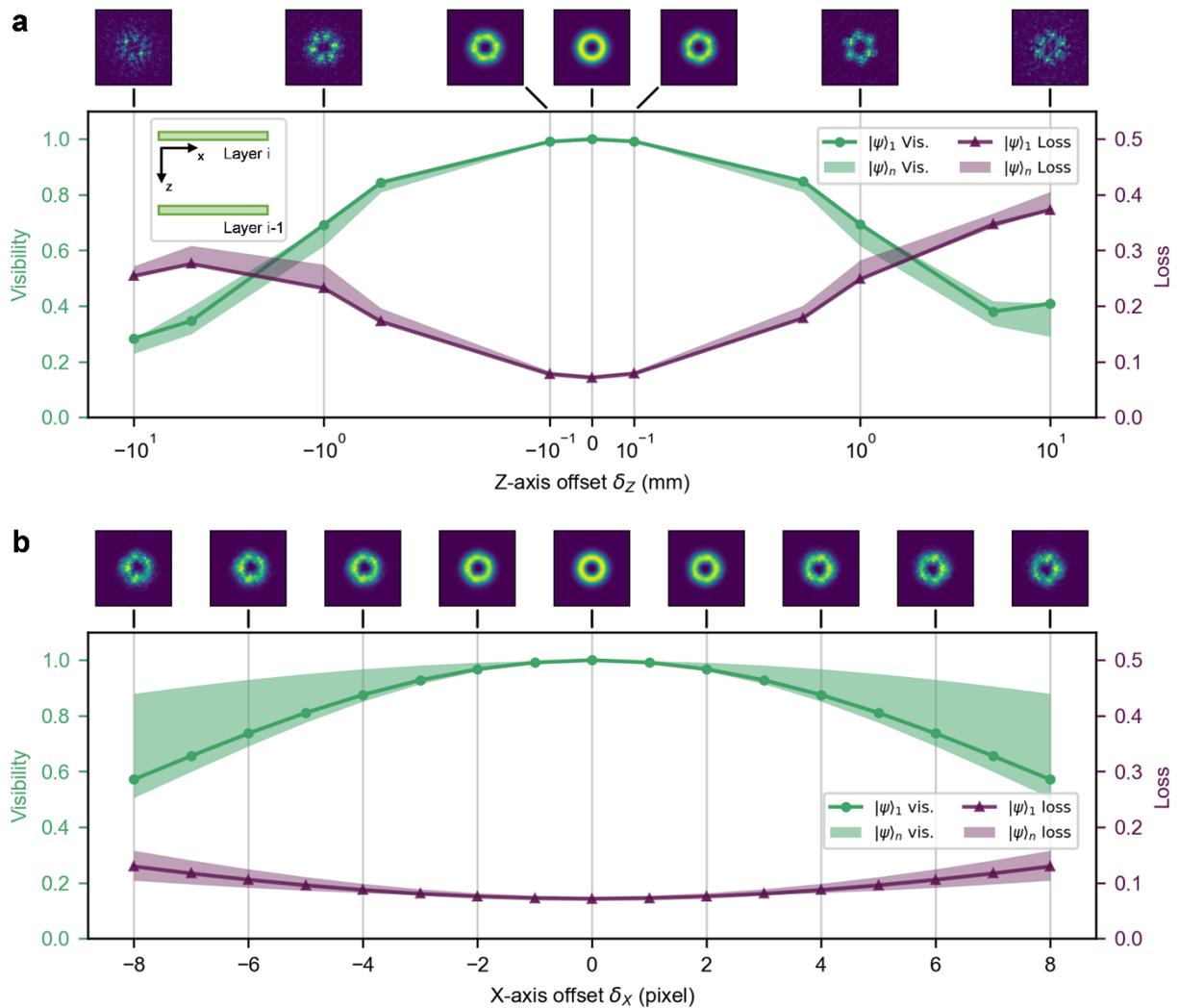

**Fig. S5. Degradation of output states due to misalignment.** The visibilities and energy losses of output states with different **a** Z-axis offsets $\delta_Z$ and **b** X-axis offsets $\delta_X$. The upper left legend in **a** provides the definition of the axes. The profiles of output state $|\psi_1\rangle$ are displayed at the top of each plot. The solid lines show the visibilities and losses of $|\psi_1\rangle$, and the filled areas represent the range of all output states $|\psi_n\rangle$.

The random x-offset follows a Gaussian distribution, and the performance of D²NN as a function of the distribution parameter $\sigma$ is shown in Fig. S6a. Larger values of $\sigma$, which indicate greater random shifts of the layers during the training process, result in deteriorated visibilities and energy losses of the output states. Fig. S6b visually illustrates the effect of different $\sigma$ values on the generated layers and corresponding output states. As $\sigma$ increases, the phase layers become smoother, leading to better tolerance to offsets but worse output state quality. These results suggest there is a trade-off between parameter $\sigma$ and the offset tolerance. Based on this analysis, we choose a value of $\sigma$ (0.3) in our demonstration that improves the smoothness of the layers while maintaining acceptable output quality.

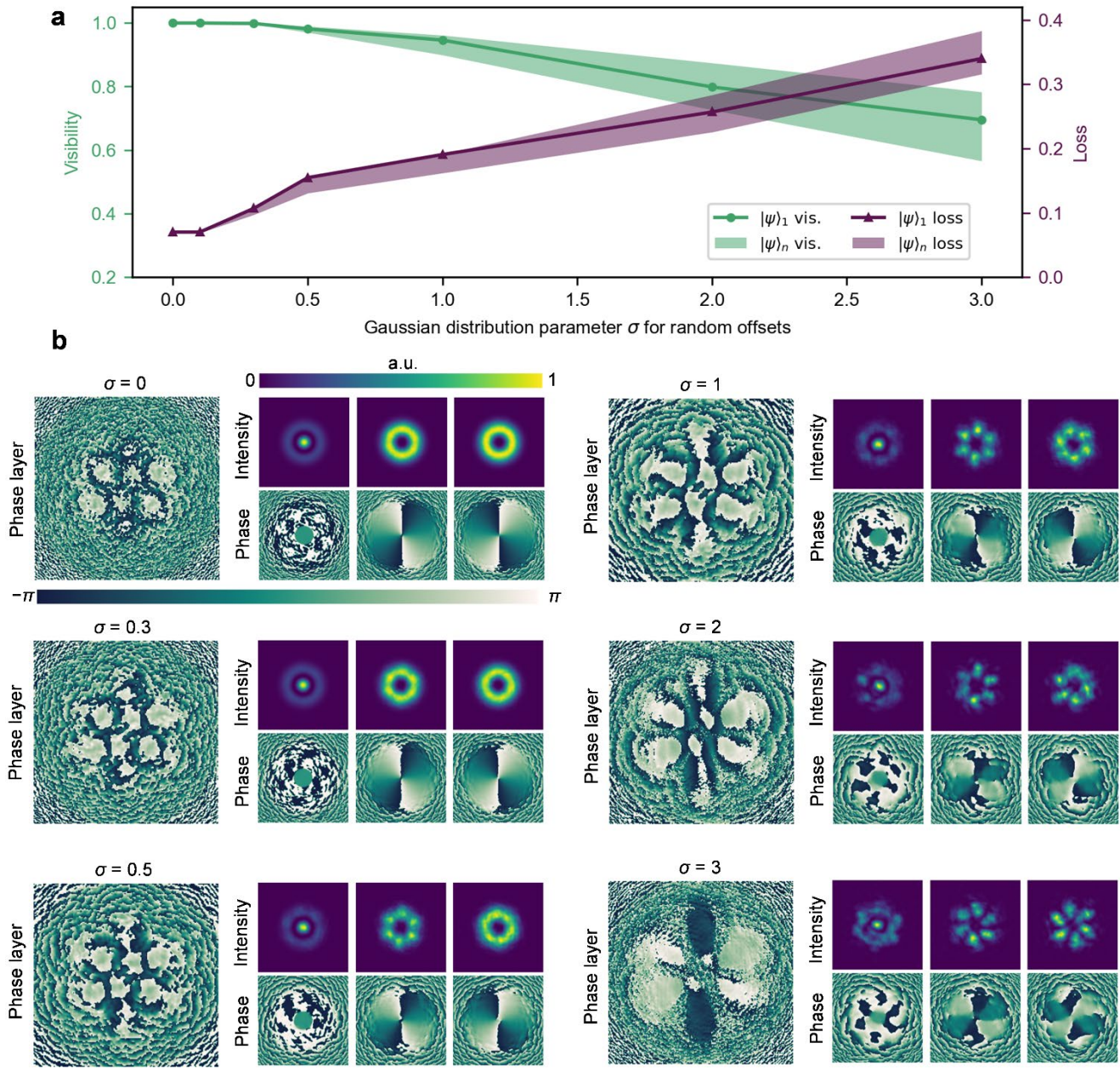

**Fig. S6. Training process with random offsets to enhance misalignment tolerance. a** The visibilities and energy losses of output states with different Gaussian distribution parameters $\sigma$ for random offsets. **b** Visualization of generated layers and output states of computational basis corresponding to different $\sigma$. Larger $\sigma$, which means higher random offset level, leads to smoother phase layers and more robustness to offsets, at the cost of the output state quality (lower visibility and higher loss).

**Supplementary Note 7: Modulation correction**

The ideal modulation of phase wraps should be as sharp as the green curve plotted in Fig. S7a. However, in reality, modulation devices are subject to finite spatial resolution, manufacturing imperfections, and fringing fields between pixels, resulting in modulation blurring at the phase wraps, known as the fringe effect. This phenomenon can introduce extra scatter and damage to performance, particularly when the phase layers of $D^2NN$ have many phase wraps as illustrated in Fig. S7b. To mitigate this degradation, one straightforward method is to include it in our $D^2NN$ model. Leveraging the flexibility of $D^2NN$, we multiply the blurring function on each layer model as a correction so that the effect of blurring is also considered during the training process. As can be seen from Fig. S7c, the generated phase layer with correction exhibits lower spatial frequencies. The phase layers used in the experiment are all after correction and have helped improve experimental performance.

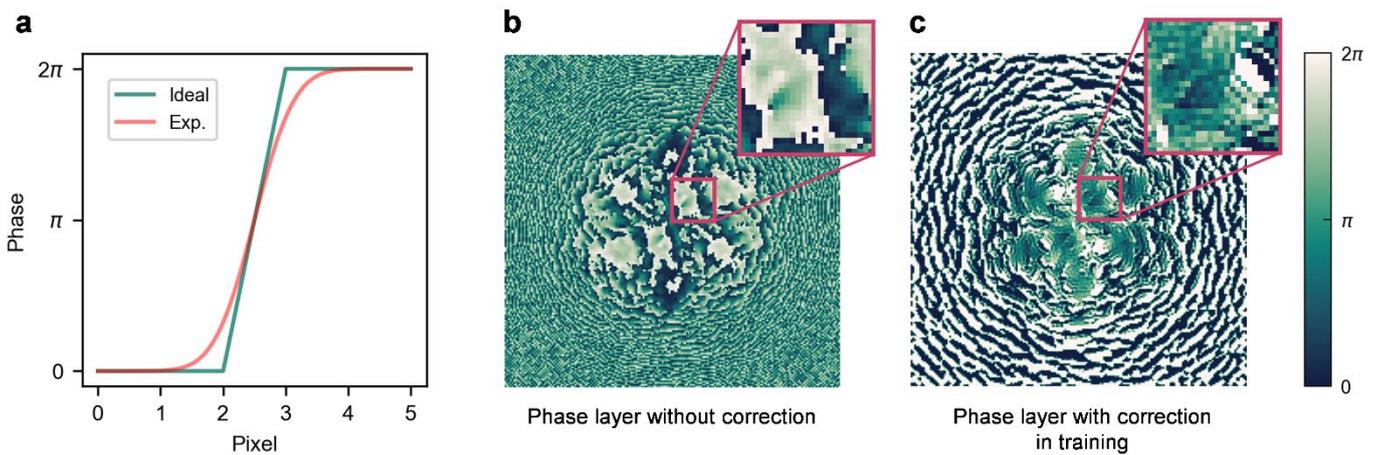

**Fig. S7. Modulation correction of the phase layers. a** The SLM modulation blurring between pixels during phase wraps of 0 and $2\pi$. **b** A typical phase layer generated without modulation correction. **c** The phase layer generated with modulation correction in the training process. In comparison, the training process with modulation correction can significantly reduce phase wraps and improve the smoothness of the phase layers.